\documentclass[11pt]{article}
\usepackage[utf8]{inputenc}
\usepackage[T1]{fontenc}
\usepackage{lmodern}
\usepackage[a4paper,margin=2.7cm]{geometry}
\usepackage[authoryear]{natbib}

\usepackage{amsmath,amssymb,amsfonts,mathtools,amsthm}
\usepackage{bm}
\usepackage{dsfont} % double-stroke indicator symbol; AG-46
\usepackage{graphicx}
\usepackage{booktabs}
\usepackage{tabularx}
\usepackage{longtable}
\usepackage{array}
\usepackage{enumitem}
\usepackage{float}
\usepackage[section]{placeins}
\usepackage{url}
\usepackage{microtype}
\usepackage{needspace} % keep the definition together with its section heading
\usepackage{xcolor}
\definecolor{revisionblue}{HTML}{000000} % muted blue for reviewed text
\DeclareRobustCommand{\rev}[1]{\textcolor{revisionblue}{#1}}
\definecolor{sarahgreen}{HTML}{000000} % Sarah Friedrich revision markings
\DeclareRobustCommand{\sfrev}[1]{\textcolor{sarahgreen}{#1}}
\definecolor{selectiongreen}{HTML}{000000} % R71, validated selection diagnostics
\DeclareRobustCommand{\selrev}[1]{\textcolor{selectiongreen}{#1}}
\usepackage{algorithm}
\usepackage{algpseudocode}
\usepackage[hidelinks]{hyperref}
\graphicspath{{./figures/}{./}}
\AtBeginDocument{\let\cite\citep}

\title{Selection instability and coefficient uncertainty in Cox modelling after Lasso selection}
\author{Lena Schemet\textsuperscript{1,*}, Andreas Groll\textsuperscript{2},
and Sarah Friedrich-Welz\textsuperscript{1,3}\\[0.8em]
\small \textsuperscript{1}Mathematical Statistics and Artificial Intelligence in Medicine,\\
\small University of Augsburg, Universit\"atsstra\ss e 14, 86159 Augsburg, Germany\\
\small \textsuperscript{2}Department of Statistics, TU Dortmund University,
44221 Dortmund, Germany\\
\small \textsuperscript{3}Center for Advanced Analytics and Predictive Sciences (CAAPS),\\
\small University of Augsburg, Universit\"atsstra\ss e 14, 86159 Augsburg, Germany\\[0.5em]
\small \textsuperscript{*}Corresponding author: \texttt{lena.schemet@uni-a.de}}
\date{}

\begin{document}
\definecolor{reviewblue}{RGB}{25,81,160}
\maketitle

\begin{abstract}
After variable selection, uncertainty quantification often treats the selected model as fixed, although it may change under repeated sampling. We investigate whether this variability can be incorporated by repeating Cox--Lasso selection within a model-based bootstrap after unpenalised Cox refitting. We compare \rev{conventional post-selection Cox Wald intervals and a model-based bootstrap without re-selection, both of which keep the selected predictor set fixed,} with two re-selection procedures: one retains the observed tuning-parameter value, whereas the other reapplies the cross-validation minimum (CV-min) tuning rule within each bootstrap sample. Against the data-generating coefficients, procedures holding the selected predictor set fixed frequently showed substantial undercoverage. Re-selection generally moved coverage closer to the nominal level, although performance varied across simulation regimes and coefficients. \rev{Sensitivity analyses evaluating coverage against the population Cox coefficients corresponding to each originally selected predictor set gave similar aggregate coverage for the primary interval comparison.} In the NKI70 breast-cancer data, the four selected genes appeared together in only 0.6\% of bootstrap samples under re-selection at the observed tuning-parameter value. These results suggest that selection instability can be an important component of post-selection coefficient uncertainty and that bootstrap re-selection can help incorporate this variability.
\end{abstract}


\section{Introduction}
\label{sec:introduction}

After variable selection, the model used for inference is itself data dependent. In time-to-event analysis, Cox--Lasso selection can determine which predictors enter a subsequent unpenalised Cox model. One analysis strategy is to select predictors using the Cox--Lasso and then refit an unpenalised Cox model using only the selected predictors \citep{tibshirani1997lasso,SchemetFriedrichWelz2026}. We refer to this set of variables as the selected predictor set. \rev{This two-stage strategy is referred to here as \emph{post-Lasso refitting}, by analogy with post-Lasso estimation in linear regression \citep{BelloniChernozhukov2013}. It is related to relaxed Lasso procedures, which allow the penalty to be reduced or removed after selection \citep{Meinshausen2007RelaxedLasso}.} Conventional standard errors and bootstrap procedures that keep this predictor set fixed quantify refitting uncertainty without repeating the selection step. If another sample selects a different set, they do not represent the additional variability arising from that change. \rev{The inferential consequences of data-dependent model selection are well recognised: conventional confidence intervals need not retain their nominal coverage after model selection \citep{Berk2013PostSelection}. One related way to account for model uncertainty is model averaging, which combines estimates across candidate models and has also been developed for Cox regression \citep{HjortClaeskens2003,HjortClaeskens2006}. Our approach addresses a different inferential question: it retains the coefficient estimate from the originally selected unpenalised Cox refit and uses bootstrap re-selection to assess uncertainty associated with the selection and refitting procedure, rather than constructing a model-averaged estimator. Bootstrap resampling has also been used to assess the stability of variable selection in Cox regression \citep{sauerbrei1992bootstrap}. However, describing selection stability does not itself establish the coverage properties of coefficient intervals. We therefore investigate how repeating Cox--Lasso selection within a model-based bootstrap affects empirical interval coverage and width.}

We use the NKI70 breast-cancer data to illustrate this distinction. The data contain 144 patients, 48 observed events, and expression measurements for 70 genes that are used as candidate predictors. \rev{10-fold} cross-validation is used with the CV-min tuning rule, which selects \(\widehat\lambda_{\min}\), the tuning-parameter value that minimises cross-validated partial-likelihood deviance. In the observed analysis, the Cox--Lasso selects four gene-expression predictors, labelled QSCN6L1, ZNF533, IGFBP5.1, and PRC1. In 500 model-based bootstrap samples in which Cox--Lasso selection is repeated at the observed tuning-parameter \rev{value~\(\widehat\lambda_{\min}\)}, this exact set reappears in 0.6\% of the samples. \rev{The same observed tuning-parameter value $\widehat\lambda_{\min}$ is used in every bootstrap sample, even though the selected predictor set can change.} Stability also differs considerably between individual predictors. PRC1 is selected again in 78.4\% of bootstrap samples, whereas QSCN6L1 is selected again in 48.6\%. These frequencies describe the particular observed-data analysis and bootstrap design; the original selected set can also depend on the random cross-validation splits. They illustrate why stability of the complete selected set and re-selection frequency for an individual predictor are distinct quantities.

The selected Cox--Lasso model also depends on how the tuning parameter is chosen. In high-dimensional Cox regression, variable-selection properties have been shown to depend strongly on the value of the Lasso tuning parameter, motivating alternative rules for choosing this value \citep{TernesRotoloMichiels2016}. This dependence is usually discussed in terms of the selected model itself, for example its sparsity or false-discovery properties. For post-selection inference, it raises a further question: whether variability introduced by tuning-parameter selection also propagates into coefficient uncertainty and confidence-interval coverage after refitting.

The inferential consequences are most apparent when selection is unstable. Reliable recovery of relevant predictors generally requires conditions on sparsity, signal strength, and predictor dependence that are difficult to assess in a single applied data set. When data perturbations change the selected set, intervals constructed as though the observed set had been prespecified may omit variation from the selection and refitting procedure. Different selected sets can also yield different coefficients for the same predictor. The implications for interval coverage therefore depend on both selection stability and the inferential target \citep{TibshiraniRinaldo2018,SchemetFriedrichWelz2026}.

Several approaches address inference after variable selection from different perspectives. Bootstrap inference after model selection has a long history \citep{efron2014modelselection}, and \citet{GarciaAnguloClaeskens2025} recently developed bootstrap inference after model selection for likelihood-based models. In Cox regression, complementary approaches include selective inference for \rev{$L_1$}-penalised likelihood models \citep{TaylorTibshirani2018}, conditional post-selection inference for interval-censored Cox data \citep{Zhang2025CoxSelective}, and debiased-Lasso procedures that target low-dimensional components of a larger Cox model \citep{XiaNanLi2023}. These approaches answer related but distinct inferential questions. Our focus is the selection and refitting procedure in which the Cox--Lasso selects predictors, an unpenalised Cox model is fitted using the selected predictors, and coefficient intervals are then reported.

The central distinction is between treating the observed selected predictor set as fixed and repeating the selection and refitting procedure. Our primary simulation question is whether the reported intervals cover the coefficients of the full data-generating Cox model, even when the selected refit omits candidate predictors. We study when holding the predictors selected in the observed data fixed omits a material component of repeated-sample coefficient variability, how repeating Cox--Lasso selection within the bootstrap changes empirical coverage and interval width, and how selection instability is reflected at the level of both the complete selected model and individual predictors. To isolate the contribution of repeated selection, we generate bootstrap samples in the same way for all procedures. The procedures differ in whether Cox--Lasso selection is repeated and whether the tuning rule is reapplied. \rev{The four primary procedures differ in how they account for uncertainty. Conventional post-selection Cox Wald intervals use the standard errors from the observed unpenalised refit, whereas the model-based bootstrap without re-selection estimates variability by repeatedly refitting the originally selected predictor set. The two re-selection procedures repeat Cox--Lasso selection followed by unpenalised refitting in each bootstrap sample. Re-selection at fixed~$\widehat\lambda_{\min}$ uses the value selected in the observed data throughout, whereas re-selection with bootstrap CV-min determines a new tuning-parameter value by cross-validation in every bootstrap sample.} Re-selection with an AIC-based tuning rule is retained only as a sensitivity analysis. Across eight simulation regimes, we interpret coverage jointly with interval width, composition of the selected predictor set, and stability diagnostics. Because aggregate coverage can mask heterogeneity across reported coefficients, we examine coverage for selected signal and noise coordinates and separately for individual coefficients. The corresponding coefficient-specific projection results are presented as a sensitivity analysis in the Supporting Material. We then use the NKI70 application to examine how model instability translates into uncertainty for individual reported coefficients.

\sfrev{We build on the model-based bootstrap of \citet{schemet2026fixedsupport}.} \rev{That study keeps the predictor set selected from the observed data fixed during bootstrap resampling. Here, we use the same bootstrap sampling procedure but repeat Cox--Lasso selection and unpenalised refitting within each bootstrap sample.} \rev{This design enables us to examine selected-set stability, individual re-selection frequencies, interval width, and coverage while using the same bootstrap sampling procedure throughout.} In the simulations, primary coverage is evaluated against the known data-generating Cox coefficients, with coverage for the selected-model projection coefficients examined in the Supporting Material. The procedures are studied as uncertainty assessments for different versions of the Cox--Lasso selection and refitting procedure, not as universal corrections for post-selection inference.

\section{Cox--Lasso selection and bootstrap re-selection}
\label{sec:methods}

This section defines the analysis procedure considered throughout the manuscript. We first describe Cox--Lasso selection and the subsequent unpenalised Cox refit, then introduce the model-based bootstrap and its re-selection extensions. Confidence-interval construction and selection-stability diagnostics are considered separately because they summarise the output of the bootstrap procedures rather than define how bootstrap samples are generated or analysed.

\subsection{Cox--Lasso selection and coefficient target}
\label{sec:target}

\rev{For subject \(i\), let \(Y_i\) denote the observed follow-up-time random variable, \(\Delta_i\) the event indicator, and \(\mathbf X_i\in\mathbb R^p\) the vector of \(p\) candidate predictors. We use \(y_i\), \(\delta_i\), and \(\mathbf x_i\) for their observed realisations.} We write
\[
\mathcal D_n
=
\rev{\{(y_i,\delta_i,\mathbf x_i):i=1,\ldots,n\}}
\]
for the observed data and consider the Cox model
\[
h(t\mid\mathbf X_i)
=
h_0(t)\exp(\mathbf X_i^\top\boldsymbol{\beta}),
\]
where \(h(t\mid\mathbf X_i)\) is the hazard at time \(t\), \(h_0(t)\) is an unspecified baseline hazard function, and \(\boldsymbol{\beta}\in\mathbb R^p\) is the regression-coefficient vector \citep{cox1972}. Conditional on the remaining predictors, \(\exp(\beta_j)\) is the hazard ratio associated with a one-unit increase in predictor \(j\). Because the baseline hazard is left unspecified, estimation of \(\boldsymbol{\beta}\) is based on the Cox partial likelihood.

Let \(\ell_n(\boldsymbol{\beta})\) denote the Cox partial log-likelihood. For a given tuning-parameter value \(\lambda\geq0\), the Cox--Lasso estimate is
\[
\widehat{\boldsymbol{\beta}}(\lambda)
=
\arg\max_{\boldsymbol{\beta}}
\left\{
\ell_n(\boldsymbol{\beta})-\lambda\|\boldsymbol{\beta}\|_1
\right\},
\qquad
\|\boldsymbol{\beta}\|_1=\sum_{j=1}^p|\beta_j|,
\]
where \(\lambda\) controls the strength of the Lasso penalty \citep{tibshirani1997lasso,simon2011regularization}. We distinguish throughout between the numerical value of the tuning parameter and the rule used to select that value. Unless stated otherwise, we use the CV-min tuning rule, which selects the tuning-parameter value minimising the cross-validated Cox partial-likelihood deviance. Writing \rev{$\Lambda$} for the candidate tuning-parameter grid and $\operatorname{CVDev}(\lambda)$ for the cross-validated deviance, this value is
\[
\widehat\lambda_{\min}
=
\underset{\lambda\in\rev{\Lambda}}{\operatorname{argmin}}
\;\operatorname{CVDev}(\lambda).
\]
\rev{The CV-min tuning rule is used in both analyses. The simulation study uses 5-fold cross-validation, whereas the NKI70 application uses 10-fold cross-validation, reflecting the different prespecified analysis settings.}

The Cox--Lasso \rev{estimate \(\widehat{\boldsymbol{\beta}}(\widehat\lambda_{\min})\), obtained from the observed data,} defines the selected predictor set 
\[
\widehat M
=
\{j:\widehat\beta_j(\widehat\lambda_{\min})\neq0\}.
\]

Having selected the predictor set \(\widehat M\), we refit an unpenalised Cox model containing only these predictors to the same data and base the subsequent coefficient inference on this refit \citep{schemet2026fixedsupport}. We denote the resulting coefficient vector by \(\widehat{\boldsymbol{\beta}}_{\widehat M}\).

Because the selected predictor set \(\widehat M\) may differ across samples, the population coefficient vector targeted by the unpenalised Cox refit \rev{likewise depends} on which predictors are included. For any candidate predictor set \(M\), let \(\boldsymbol{\beta}_M\) denote the population coefficient vector targeted by the \rev{unpenalised} Cox partial-likelihood fit restricted to the predictors in \(M\). For the simulation study, our primary evaluation asks whether an interval reported for an originally selected \rev{predictor} covers the corresponding data-generating coefficient \(\beta_j^\circ\). We also consider coverage of the \rev{Cox projection for the selected model} \((\boldsymbol{\beta}_{\widehat M})_j\) as a sensitivity analysis, because this is the population coefficient targeted by the selected unpenalised refit.
\rev{A Cox model fitted using only the selected predictors can target a different population coefficient for predictor \(j\) than the full data-generating Cox model if other prognostic predictors are omitted. Consequently, \((\boldsymbol{\beta}_{\widehat M})_j\) need not equal \(\beta_j^\circ\), even though both refer to predictor \(j\). Predictor correlation may further affect this discrepancy.}
\sfrev{In the two-predictor example in Supporting Material, Section~S2.3, the data-generating coefficient for $X_1$ is $0.8$, while the estimated Cox projection after omitting $X_2$ is $0.947$.}

\Needspace{9\baselineskip}
\subsection{Model-based bootstrap and re-selection}
\label{sec:bootstrap}

\rev{We take as our starting point the \emph{model-based bootstrap} introduced by \citet{schemet2026fixedsupport}. Throughout this manuscript, this name refers to the procedure \emph{without re-selection}: Cox--Lasso first selects the predictor set \(\widehat M\) in the observed data, and an unpenalised Cox model is refitted using this set. The same predictors are retained in every bootstrap sample, where only the unpenalised Cox model is refitted. The resulting bootstrap distribution thus reflects coefficient variability when the originally selected predictor set is treated as fixed.}

\rev{To generate the bootstrap samples, the cumulative baseline hazard is estimated from the observed-data Cox refit using the Breslow estimator, while the censoring distribution is estimated using a reverse Kaplan--Meier plug-in estimator. The observed covariate vectors are held fixed. Bootstrap event times are drawn from the fitted Cox time-to-event distribution, and censoring times independently from the estimated censoring distribution.}

\rev{Our re-selection procedures generate bootstrap samples in the same way, but repeat Cox--Lasso selection before unpenalised refitting, allowing the selected predictor set to change.}

Re-selection can be implemented in different ways depending on how the tuning parameter is handled across bootstrap samples. \rev{For the primary comparison, we distinguish two re-selection strategies: \emph{re-selection at fixed~\(\widehat\lambda_{\min}\)} and \emph{re-selection with bootstrap CV-min tuning}.} Under re-selection at \rev{fixed~\(\widehat\lambda_{\min}\)}, the Cox--Lasso is fitted in every bootstrap sample using the tuning-parameter value selected in the observed data. The tuning-parameter value therefore remains fixed while the selected predictor set is allowed to change. Alternatively, the tuning rule itself can be reapplied within each bootstrap sample, so that both the selected tuning-parameter value and the selected predictor set may vary. In this case, the same rule as in the original analysis, such as the CV-min rule, or another prespecified tuning criterion, such as an AIC-based rule, can be used. \rev{We refer to the latter sensitivity procedure as \emph{re-selection with bootstrap AIC tuning}. Algorithm~\ref{alg:reselection-bootstrap} uses the shorter labels \emph{fixed~\(\widehat\lambda_{\min}\)}, \emph{bootstrap CV-min}, and \emph{bootstrap AIC} for these three re-selection strategies.}

\rev{We use the superscript \((b)\) to index bootstrap replicates, with \(b=1,\ldots,B\).} We write
\(\mathcal D_n^{\rev{(b)}}\) for bootstrap data set \(b\),
\(\widehat\lambda^{\rev{(b)}}\) for the tuning-parameter value
used in that bootstrap sample, and
\(\widehat M^{\rev{(b)}}\) for the resulting bootstrap-selected
predictor set.

For each bootstrap sample, we store the coefficients
from the unpenalised Cox refit in a vector
\(\widehat{\boldsymbol{\beta}}^{\rev{(b)}}\in\mathbb R^p\),
with one coordinate for each original candidate predictor.
If predictor \(j\) is selected, its coordinate contains
the corresponding refitted coefficient; otherwise,
it is set to zero. This representation aligns the
coefficients across bootstrap samples with different
selected predictor sets. A stored zero indicates
non-selection rather than an estimated zero effect. \rev{Algorithm~\ref{alg:reselection-bootstrap} summarises the procedure.}

\begin{algorithm}[t]
\caption{Model-based bootstrap with Cox--Lasso re-selection}
\label{alg:reselection-bootstrap}
\begin{algorithmic}[1]
\Require Observed data \(\mathcal D_n\) and number of bootstrap replicates \(B\)
\Require Re-selection strategy \(s\in\{\text{fixed }\widehat\lambda_{\min},\text{bootstrap CV-min},\text{bootstrap AIC}\}\)
\Ensure \selrev{Bootstrap confidence intervals and selection-stability measures}
\State Apply the CV-min tuning rule to the observed data to obtain \(\widehat\lambda_{\min}\)
\State Fit the Cox--Lasso at \(\widehat\lambda_{\min}\) and obtain the selected predictor set \(\widehat M\)
\State Refit an unpenalised Cox model using the selected predictor set \(\widehat M\)
\State \rev{From the observed data, estimate the cumulative baseline hazard using the unpenalised Cox refit and the censoring distribution using reverse Kaplan--Meier}
\For{\(b=1,\ldots,B\)}
    \State Generate a bootstrap data set \(\mathcal D_n^{\rev{(b)}}\) from the fitted time-to-event and censoring distributions
    \If{\(s=\text{fixed }\widehat\lambda_{\min}\)}
        \State \(\widehat\lambda^{\rev{(b)}}\gets\widehat\lambda_{\min}\)
    \ElsIf{\(s=\text{bootstrap CV-min}\)}
        \State Apply the CV-min tuning rule to \(\mathcal D_n^{\rev{(b)}}\) to obtain \(\widehat\lambda^{\rev{(b)}}\)
    \Else
        \State Apply the AIC-based tuning rule to obtain \(\widehat\lambda^{\rev{(b)}}\)
    \EndIf
    \State Fit the Cox--Lasso at \(\widehat\lambda^{\rev{(b)}}\) and obtain the bootstrap-selected predictor set \(\widehat M^{\rev{(b)}}\)
    
    \State Initialise \(\widehat{\boldsymbol{\beta}}^{\rev{(b)}}
    \gets \boldsymbol{0}_p\)
    
    \If{\(\widehat M^{\rev{(b)}}\neq\varnothing\)}
        \State Refit an unpenalised Cox model using the bootstrap-selected predictor set \(\widehat M^{\rev{(b)}}\)
        \State Store the refitted coefficients in their corresponding coordinates of \(\widehat{\boldsymbol{\beta}}^{\rev{(b)}}\)
    \EndIf
\EndFor
\State \selrev{Calculate bootstrap confidence intervals and selection-stability measures (Sections~\ref{sec:ci-construction} and~\ref{sec:stability})}
\end{algorithmic}
\end{algorithm}

Formal inferential theory for this particular combination of
repeated Cox--Lasso selection, unpenalised refitting,
and model-based bootstrap resampling is, to our knowledge,
not yet available. Its development is challenging because
the selection step is non-smooth and data dependent,
while re-selection allows both the selected predictor set
and, when the tuning rule is reapplied, the tuning-parameter
value to vary across bootstrap samples.

Rather than imposing additional theoretical assumptions,
we therefore begin by studying the finite-sample behaviour
of the proposed procedures empirically. Our aim is to assess
whether allowing the selection step to vary captures a
relevant component of coefficient uncertainty.

The results should therefore be interpreted as an empirical
evaluation rather than as a general validity result for
post-selection inference. Formal analysis of the asymptotic
properties and inferential guarantees remains for future work.

\subsection{Confidence-interval construction}
\label{sec:ci-construction}

We distinguish the conventional Cox Wald interval obtained after the observed unpenalised refit \rev{\citep{therneau2000modeling}} from confidence intervals constructed using the bootstrap distributions defined above. For a predictor \(j\in\widehat M\) selected in the observed data, let \(\widehat{\mathrm{se}}_{j,\widehat M}\) denote the conventional standard error \rev{computed from the unpenalised Cox refit on the observed data using the selected predictor set \(\widehat M\)}. At nominal level \(1-\alpha\), the corresponding post-selection Cox Wald interval is
\begin{equation}
\left[
\widehat\beta_{j,\widehat M}
-
z_{1-\alpha/2}\widehat{\mathrm{se}}_{j,\widehat M},
\;
\widehat\beta_{j,\widehat M}
+
z_{1-\alpha/2}\widehat{\mathrm{se}}_{j,\widehat M}
\right],
\label{eq:postselection-wald}
\end{equation}
where \(z_{1-\alpha/2}\) denotes the corresponding standard-normal quantile. 

This conventional Cox Wald interval treats the selected
predictor set as if it had been specified in advance.
\rev{No bootstrap resampling is required to construct this interval.}
Its standard error does not account for the preceding
Cox--Lasso selection step, and nominal coverage is
therefore not generally guaranteed
\citep{Berk2013PostSelection,TaylorTibshirani2018}.

For a given bootstrap procedure, the variability of predictor \(j\) across the \(B\) bootstrap replicates can be summarised by
\[
\widehat{\mathrm{se}}_{j,\mathrm{boot}}
=
\operatorname{sd}\!\left(
\widehat\beta_j^{\rev{(1)}},\ldots,\widehat\beta_j^{\rev{(B)}}
\right),
\]
where the \rev{bootstrap coefficients} are defined in Section~\ref{sec:bootstrap}. Replacing the conventional standard error \rev{by the corresponding empirical standard error computed from the bootstrap replicates} gives the bootstrap-SE Wald interval
\begin{equation}
\left[
\widehat\beta_{j,\widehat M}
-
z_{1-\alpha/2}\widehat{\mathrm{se}}_{j,\mathrm{boot}},
\;
\widehat\beta_{j,\widehat M}
+
z_{1-\alpha/2}\widehat{\mathrm{se}}_{j,\mathrm{boot}}
\right].
\label{eq:bootstrap-wald-main}
\end{equation}
This construction corresponds to a standard-normal bootstrap interval based on a bootstrap estimate of the standard error \citep{efronTibshirani1993,davisonHinkley1997}. Its centre is the coefficient from the observed unpenalised Cox refit, while its width is determined by the variability represented by the corresponding bootstrap distribution. When the selected predictor set is held fixed, this variability reflects repeated unpenalised refitting of the predictor set selected in the observed data. Under re-selection, the bootstrap distribution may additionally reflect changes in the selected predictor set and, when the tuning rule is reapplied, changes in the tuning-parameter value. \rev{For a predictor selected in the observed data, non-selection in a bootstrap sample results in a stored zero, as defined in Section~\ref{sec:bootstrap}. These zeros are included when computing \(\widehat{\mathrm{se}}_{j,\mathrm{boot}}\); they are not treated as missing values or omitted.}

Confidence intervals can also be constructed directly from the empirical bootstrap distribution \citep{efronTibshirani1993,davisonHinkley1997}. Let \(\widehat Q^*_{j,\gamma}\) denote the empirical \(\gamma\)-quantile of \(\widehat\beta_j^{\rev{(1)}},\ldots,\widehat\beta_j^{\rev{(B)}}\). The percentile interval is
\[
\left[
\widehat Q^*_{j,\alpha/2},
\widehat Q^*_{j,1-\alpha/2}
\right],
\]
whereas the reverse-percentile basic interval is
\[
\left[
2\widehat\beta_{j,\widehat M}
-
\widehat Q^*_{j,1-\alpha/2},
\;
2\widehat\beta_{j,\widehat M}
-
\widehat Q^*_{j,\alpha/2}
\right].
\]
The percentile interval uses the empirical bootstrap quantiles directly, whereas the reverse-percentile basic interval reflects these quantiles about the observed coefficient estimate. Thus, the bootstrap-SE Wald, percentile, and reverse-percentile basic constructions use the same bootstrap coefficient distribution but extract uncertainty from it in different ways. \rev{Under re-selection, this distribution can contain a point mass at zero and may be asymmetric. No additional adjustment for this feature is made: non-selection zeros are retained in the empirical quantiles used for the percentile and reverse-percentile basic intervals. The bootstrap-SE Wald interval remains symmetric by construction and may not reflect such asymmetry. None of these constructions is guaranteed to achieve nominal post-selection coverage, and we assess their behaviour empirically.}

These interval constructions specify how the bootstrap distributions are translated into confidence intervals. Their use in the present setting does not by itself establish general post-selection validity. In particular, the bootstrap-SE Wald construction should not be interpreted as implying a general Wald approximation after variable selection \citep{efron2014modelselection,TaylorTibshirani2018,TibshiraniRinaldo2018}. The procedures are therefore defined here as alternative ways of summarising the uncertainty represented by the corresponding bootstrap distributions. Their finite-sample properties are investigated empirically in the subsequent simulation study. \rev{Definitions of the simulation estimands and stability summaries are given in Supporting Material, Section~S1. Bootstrap implementation details are provided in Section~S2.}

\subsection{Selection-stability diagnostics}
\label{sec:stability}

The bootstrap distributions under re-selection contain information not only about coefficient variability but also about how reproducibly the selected model is recovered. Bootstrap resampling has long been used to assess the stability of variable-selection procedures, including in Cox regression \citep{sauerbrei1992bootstrap}. We therefore distinguish coordinate-level stability from stability of the selected predictor set.

For a predictor \(j\in\widehat M\) selected in the observed data, coordinate-level stability is described by its re-selection frequency, that is, the proportion of bootstrap samples for which \(j\in\widehat M^{\rev{(b)}}\). As described in Section~\ref{sec:bootstrap}, we record a value of zero for predictor \(j\) in bootstrap samples in which it is not selected. The corresponding zero frequency is therefore the proportion of bootstrap samples in which \(j\) is not selected. Hence, for an originally selected predictor, the zero frequency equals one minus its re-selection frequency, and the two quantities are equivalent summaries of coordinate-level selection behaviour.

Stability of the selected predictor set concerns the selected model as a whole. We consider the proportion of bootstrap samples that exactly reproduce the observed selected predictor set, the number of predictors selected in each bootstrap sample, and \rev{the Jaccard index. For the observed selected predictor set \(\widehat M\) and a bootstrap-selected set \(\widehat M^{(b)}\), the index for each bootstrap sample is defined as}
\[
\frac{
|\widehat M\cap\widehat M^{\rev{(b)}}|
}{
|\widehat M\cup\widehat M^{\rev{(b)}}|
}.
\]
The Jaccard index is a standard measure of similarity between sets \citep{jaccard1901} and has also been used to quantify the stability of selected feature sets \citep{kalousis2007stability,nogueira2018stability}. It equals one when the two selected predictor sets coincide and decreases as their overlap becomes smaller. \rev{Exact-set reproduction indicates whether the complete originally reported predictor set reappears, whereas the Jaccard index also records partial overlap when only some predictors recur. Both measure reproducibility of the observed selection under bootstrap resampling, not its correctness: a frequently reproduced set may still omit relevant predictors or contain noise predictors.}

These quantities describe the stability of the
repeated-selection procedure, following the use of
bootstrap selection frequencies and set-overlap measures
to assess variable-selection stability
\citep{sauerbrei1992bootstrap,nogueira2018stability}.
Considering these diagnostics alongside the bootstrap
coefficient distributions helps distinguish settings
in which the selected predictor set remains relatively
stable from those in which substantial changes in
selection accompany coefficient variability.
These diagnostics are descriptive and do not establish
inferential validity. \rev{Selected-set size is a descriptive count, not a comparison with a reference set.} Formal definitions are provided
in Supporting Material, Section~S1.

\section{Simulation study}
\label{sec:simulation-plan}

The NKI70 data example illustrates that Cox--Lasso selection can be unstable, but a single data set cannot establish repeated-sampling behaviour. The simulation study therefore evaluates whether allowing the Cox--Lasso selection step to vary within the model-based bootstrap changes uncertainty assessment relative to procedures that treat the selected predictor set as fixed. We additionally investigate how this behaviour depends on candidate-set size, predictor correlation and substitution, signal strength, censoring, and reapplication of the CV-min tuning rule within bootstrap samples.

\subsection{Design}

The main comparison consists of (i) post-selection Cox Wald
intervals after the unpenalised Cox refit on the selected
predictor set, (ii) model-based bootstrap intervals,
(iii) re-selection at \rev{fixed~\(\widehat\lambda_{\min}\)},
and (iv) re-selection with the CV-min tuning rule reapplied
in each bootstrap sample. All three bootstrap procedures
based on the model-based bootstrap data-generating
mechanism use the bootstrap-SE Wald interval from
Section~\ref{sec:ci-construction}. Re-selection with an
AIC-based tuning rule and alternative interval
constructions are examined separately as sensitivity
analyses. The primary evaluation uses the known coefficients of the data-generating Cox model. \rev{Their nonzero entries define the known active set \(M_0\), which allows selected predictors to be classified as signals or noise.} This choice affects how we assess the intervals, not how they are constructed. \sfrev{Because $M_0$ is known, we can compare the original and bootstrap selections with the true active set and assess agreement with the original selection separately.}

Table~\ref{tab:sim-regimes-main} summarises eight
simulation regimes, each crossed with
\(n\in\{75,150,300,600\}\), giving 32 settings.
Each setting uses 900 Monte Carlo replicates,
\(B=500\) bootstrap replicates per Monte Carlo
replicate, and nominal 90\% intervals.
\rev{Given the substantial computational cost of the nested bootstrap procedures, we chose 900 Monte Carlo replicates as the lowest number we judged adequate for identifying meaningful departures from nominal coverage. For a simple binomial coverage proportion of 0.90, 900 independent replicates yield a \sfrev{Monte Carlo standard error (MCSE)} of approximately 0.010 (one percentage point). This is a benchmark rather than the exact MCSE of pooled selected-coordinate coverage (Supporting Material, Section~S2), and small differences between procedures may remain unresolved.}
In each Monte Carlo sample, \rev{5-fold cross-validation}
is used with the CV-min tuning rule to obtain the
tuning-parameter \rev{value~\(\widehat\lambda_{\min}\)}
and the corresponding selected predictor set.
The grid is deliberately structured rather than fully
factorial: it increases the candidate-set size from
\(p=10\) to \(p=50\), introduces correlation and
competing predictors, weakens signals, and reduces
event information through heavier censoring.

\rev{Raw covariates are generated independently as \(\mathbf Z_i\sim\mathcal N_p(\mathbf 0,\boldsymbol\Sigma)\) with \(\Sigma_{jk}=\rho^{|j-k|}\), an AR(1)-type correlation structure across predictors rather than over time. Within each Monte Carlo replicate, covariates are centred and divided by their empirical standard deviations to yield \(\mathbf X_i\). Thus, the data-generating coefficients \(\boldsymbol\beta^\circ\) refer to standardised predictors.}

The coefficient pattern used in each regime is indicated
in Table~\ref{tab:sim-regimes-main}. The standard four-signal pattern is
\((\beta_1^\circ,\ldots,\beta_4^\circ)
=(0.8,0.7,0.5,0.8)\);
the weak-signal pattern is
\((0.45,0.35,0.25,0.45)\);
and the \rev{proxy/competing-predictor pattern} has
\(\beta_1^\circ=0.8\) and \(\beta_6^\circ=0.5\).
All remaining coefficients are zero.
\rev{We refer to this as the proxy/competing-predictor regime because, with \(\rho=0.7\), inactive predictors neighbouring the two signals may act as proxies and compete for inclusion during Cox--Lasso selection.}

For each subject, event times \(T_i\) are generated
under a proportional-hazards model with linear
predictor \(\mathbf X_i^\top\boldsymbol{\beta}^{\circ}\)
and a Weibull baseline distribution with shape 2
and scale 1. We use inverse-transform
sampling for the Cox model as described by \citet{bender2005}.
Independent right censoring is introduced through
exponentially distributed censoring \rev{times~\(C_i\)}.
The exponential rate is calibrated to obtain the
prespecified marginal censoring proportion in each
simulation regime, following the general
censoring-calibration approach of \citet{ramos2024}.
The observed time-to-event outcome is
\(Y_i=\min(T_i,C_i)\), with event indicator
\(\Delta_i=\rev{\mathds{1}\{T_i\leq C_i\}}\).
Further details of the censoring calibration and
implementation are provided in the Supporting
Material, Section~S2 (Simulation protocol and
implementation).

\begin{table}[ht]
\centering
\caption{
Main simulation regimes, each evaluated for
\(n\in\{75,150,300,600\}\).
The coefficient patterns are denoted by S
(standard four-signal), W (weak four-signal),
and P (two-signal proxy), with nonzero
coefficients specified in the text.
}
\small
\begin{tabular}{
p{0.37\linewidth}
p{0.10\linewidth}
p{0.10\linewidth}
p{0.15\linewidth}
p{0.11\linewidth}
}
\toprule
\textbf{Regime} &
\textbf{\(p\)} &
\textbf{\(\rho\)} &
\textbf{Censoring} &
\textbf{Pattern} \\
\midrule
Low-dimensional baseline
    & 10 & 0   & 10\% & S \\
Moderate candidate set
    & 25 & 0.3 & 10\% & S \\
Correlated predictors
    & 25 & 0.5 & 10\% & S \\
Larger candidate set
    & 50 & 0   & 10\% & S \\
Larger candidate set, correlated
    & 50 & 0.5 & 10\% & S \\
\rev{Weak signals (correlated)}
    & 50 & 0.5 & 10\% & W \\
Heavy censoring
    & 50 & 0.5 & 40\% & S \\
Proxy/competing predictors
    & 50 & 0.7 & 10\% & P \\
\bottomrule
\end{tabular}
\label{tab:sim-regimes-main}
\end{table}
\FloatBarrier

\subsection{Evaluation criteria}
\label{sec:performance}

We evaluate confidence intervals for coefficients selected
in each original Monte Carlo sample. 
\rev{All simulation intervals have nominal coverage of 90\% ($\alpha=0.10$).} Let
\(\widehat M^{(r)}\) denote the predictor set selected in
replicate \(r\). Within each setting, \emph{selected-coordinate
coverage} is the proportion of numerically available
replicate--coordinate pairs \((r,j)\), with
\(j\in\widehat M^{(r)}\), whose interval contains the data-generating coefficient \(\beta_j^\circ\). This summary concerns originally selected coordinates and is not conditional on one fixed selected model \citep{kammer2022}. Because an unpenalised refit with selected predictors may target a different population coefficient, we also evaluate coverage of the corresponding \rev{Cox projection for the selected model} \((\boldsymbol{\beta}_{\widehat M^{(r)}})_j\) in the Supporting Material, Section~S3.5. \rev{Thus, the same reported interval for each predictor selected in the original Monte Carlo sample is assessed against two different coefficient targets. Predictors not selected in that original sample do not enter either coverage summary. }

\rev{The primary interval-performance measures are selected-coordinate coverage and mean interval width for the four main procedures.} In addition to coverage and mean interval width, we examine coefficient-specific coverage among Monte Carlo replicates in which each predictor was originally selected. We also report coverage stratified by nonzero versus zero data-generating coefficients among originally selected coordinates. This signal/noise classification refers to \(\beta_j^\circ\), not the Cox projection, and is not interpreted as a type-I-error assessment. \rev{The Cox projection target, alternative interval constructions, and bootstrap AIC tuning are sensitivity analyses; selection-recovery and re-selection counts are descriptive diagnostics rather than measures of interval validity.}

\selrev{We compare the original Cox--Lasso selection with the known active set $M_0$. Exact-set recovery is the proportion of Monte Carlo replicates in which $\widehat M^{(r)}=M_0$, while the Jaccard index measures partial overlap. For the bootstrap procedures, we calculate expected numbers of true positives (TP), false positives (FP), and false negatives (FN) from the recorded selection frequencies. We also calculate the expected Hamming distance, the number of predictors on which a bootstrap selection differs from either $M_0$ or the original selection $\widehat M^{(r)}$. Results are averaged within each setting and then equally across the 32 settings.}

\selrev{The selection frequencies show how often each predictor is selected, but not which predictors are selected together in each bootstrap sample. We therefore cannot calculate bootstrap exact-set agreement or mean Jaccard overlap from the stored simulation results. Both measures are available for NKI70, where the complete bootstrap selections were retained. Definitions are given in Supporting Material, Section~S1.}

We also report
\(\lvert(\boldsymbol{\beta}_{\widehat M^{(r)}})_j
-\beta_j^\circ\rvert\)
to describe target differences.

These diagnostics complement the coverage analysis but
are not interpreted as criteria for inferential validity.

Operating characteristics are first calculated within
each of the 32 simulation settings, excluding and
separately tracking numerically unavailable intervals.
Reported overall means give equal weight to the 32
setting-specific estimates and are interpreted alongside
setting-specific and coefficient-specific results
\citep{morris2019}. \selrev{Additional coverage results and sensitivities are reported in Supporting Material, Section~S3.}

\subsection{Results}

\rev{We first summarise coverage against the data-generating coefficients across all 32 simulation settings in Table~\ref{tab:main-wald-comparison}.} We then examine the selected predictor sets and coverage for individual coefficients. Analyses using the Cox projection targets and alternative interval constructions are reported in the Supporting Material.

\subsubsection{Overall performance}

Holding the selected predictor set fixed resulted in undercoverage across a broad range of settings. Coverage was \rev{below~0.80} in 29 of 32 settings for post-selection Cox Wald intervals and in 28 settings for the model-based bootstrap. Under re-selection at \rev{fixed~\(\widehat\lambda_{\min}\)}, no setting had coverage \rev{below~0.80}, whereas under re-selection with bootstrap CV-min tuning, one setting did. The equal-weighted mean selected-coordinate coverages against the data-generating coefficients were 0.680, 0.721, 0.866, and 0.888, respectively.

These means are used to place the methods in an overall context rather than to characterise every selected coefficient. An average close to the nominal level can mask undercoverage for some selected coefficients and overcoverage for others. \rev{Among the four primary procedures, re-selection with bootstrap CV-min tuning had mean coverage closest to 0.90, but also wider intervals than re-selection at fixed~\(\widehat\lambda_{\min}\). This descriptive result does not establish it as uniformly preferable across settings or guarantee post-selection validity.} We therefore interpret these summaries together with the setting-specific, composition-specific, and coefficient-level results.

\begin{table}[ht]
\centering
\caption{Overall comparison across the 32 simulation settings at nominal 90\% coverage.
``Settings $<0.80$'' denotes the number of settings with selected-coordinate coverage \rev{below~0.80}.
Mean absolute deviation summarises deviations of setting-specific coverage from 0.90. Coverage is evaluated against the data-generating coefficients \(\beta_j^\circ\). \rev{Bold entries indicate the fewest settings below 0.80, mean coverage closest to 0.90, smallest mean absolute deviation, and narrowest mean interval, respectively. These column-wise extrema do not identify an overall best procedure.}}
\label{tab:main-wald-comparison}
\small
\setlength{\tabcolsep}{5pt}
\resizebox{\linewidth}{!}{%
\begin{tabular}{lrrrr}
\toprule
\textbf{Method} & \textbf{Settings $<0.80$} & \textbf{Mean coverage} & \textbf{Mean abs. deviation} & \textbf{Mean width} \\
\midrule
Post-selection Cox Wald & 29 & 0.680 & 0.220 & \rev{\textbf{0.344}} \\
Model-based bootstrap & 28 & 0.721 & 0.179 & 0.372 \\
Re-selection, \rev{fixed~$\widehat\lambda_{\min}$} & \rev{\textbf{0}} & 0.866 & 0.035 & 0.428 \\
Re-selection, bootstrap CV-min & 1 & \rev{\textbf{0.888}} & \rev{\textbf{0.033}} & 0.492 \\
\bottomrule
\end{tabular}%
}
\end{table}

The differences in coverage were accompanied by wider intervals. Mean width increased from 0.372 for the model-based bootstrap to 0.428 under re-selection at \rev{fixed~\(\widehat\lambda_{\min}\)} and 0.492 under re-selection with bootstrap CV-min tuning. The empirical comparison therefore concerns both coverage and the additional uncertainty reflected in interval width. It should not be interpreted as evidence of uniformly nominal coverage under re-selection.

\begin{figure}[htbp]
\centering
\includegraphics[width=0.98\linewidth]{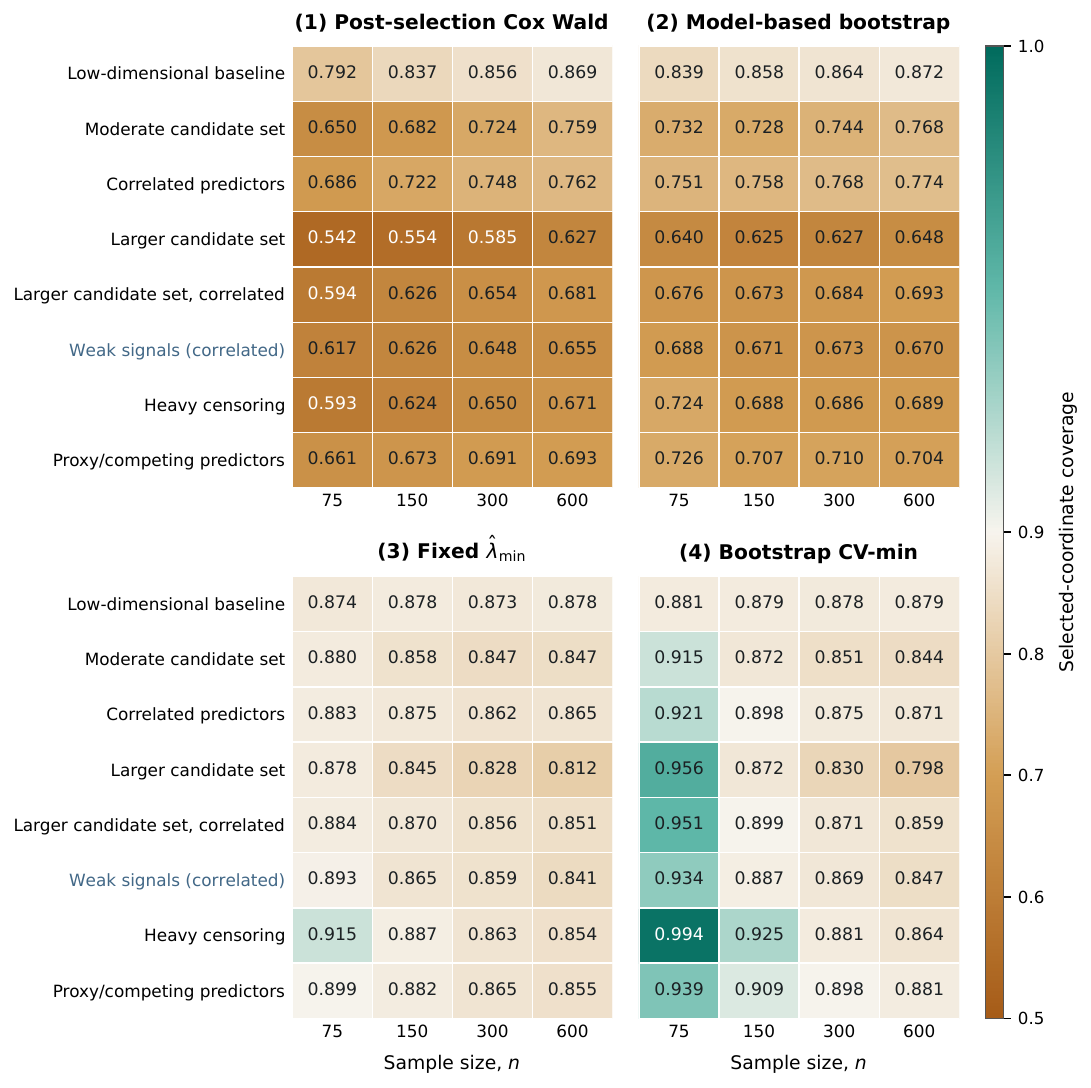}
\caption{Selected-coordinate coverage for the four primary procedures across the 32 simulation settings (nominal level 0.90). Panels show (1) post-selection Cox Wald, (2) model-based bootstrap without re-selection, (3) re-selection at \rev{fixed~$\widehat\lambda_{\min}$}, and (4) re-selection with bootstrap CV-min tuning. Bootstrap procedures use bootstrap-SE Wald intervals. Each cell represents one regime and sample size; all panels use the same colour scale. Coverage is evaluated against the data-generating coefficients for originally selected coordinates. \rev{Orange tones indicate lower coverage and green tones higher coverage. The target is 0.90 rather than the darkest green: values below 0.90 indicate undercoverage, while values above 0.90 indicate overcoverage.}}
\label{fig:four-method-coverage-main}
\end{figure}

Performance nevertheless varied substantially across regimes. In the low-dimensional baseline, differences between the model-based bootstrap and the re-selection procedures were comparatively modest, and all methods moved closer together as \(n\) increased. In contrast, in the larger-candidate-set regimes the gap between approaches was substantial: post-selection Cox Wald intervals and the model-based bootstrap often showed clear undercoverage, whereas the two re-selection procedures were much closer to the nominal level, especially at small and moderate sample sizes. Figure~\ref{fig:four-method-coverage-main} compares selected-coordinate coverage for all four primary procedures across the 32 settings, using \(\beta_j^\circ\) as the coverage target. \rev{For nominal 90\% intervals, coverage closer to 0.90 is preferable to either undercoverage or overcoverage, but coverage must also be considered alongside interval width. The AIC-based re-selection strategy is omitted from this primary four-method coverage figure; it is retained as a sensitivity in Figure~\ref{fig:coverage-width-tradeoff} and the Supporting Material.} Full setting-specific coverage and interval widths are reported in Supporting Material, Section~S3.

\subsubsection{Performance by composition of the selected predictor set}

The decrease in selected-coordinate coverage with increasing sample size in several larger candidate-set regimes should not be interpreted as a coverage curve for one fixed coefficient. The collection of coordinates entering the summary changes because the selected predictor set changes across Monte Carlo samples. Consequently, changes in selected-coordinate coverage can reflect both changes in coefficient-specific performance and changes in the composition of the selected predictor set.

Table~\ref{tab:composition-transition} summarises this transition from \(n=75\) to \(n=600\). Predictors with nonzero data-generating coefficients were selected with high probability in most regimes, but the mean number of additionally selected predictors with zero data-generating coefficients increased with sample size. Coverage among these selected noise coordinates was generally lower than coverage among selected signal coordinates. \rev{For all three coverage columns in Table~\ref{tab:composition-transition}, the nominal benchmark is 0.90, not the largest observed coverage value.} \rev{The noise counts are means per original Monte Carlo sample, not percentages or totals across the Monte Carlo replicates. Although larger samples might be expected to improve variable selection, the mean number of selected noise predictors increased in these regimes. This is compatible with the prediction-oriented nature of CV-min tuning, which does not aim at exact recovery of the active set. More precise estimation at larger sample sizes does not necessarily imply fewer false-positive selections.}

\begin{table}[ht]
\centering
\caption{Original-model composition and selected-coordinate coverage under bootstrap CV-min re-selection. \rev{Arrows compare \(n=75\) with \(n=600\). ``Selected noise'' is the mean number of originally selected predictors with zero data-generating coefficients; coverage is shown for selected signal, noise and all coordinates (nominal level 0.90).}}
\label{tab:composition-transition}
\footnotesize
\setlength{\tabcolsep}{5pt}
\begin{tabular}{@{}l@{\hspace{10pt}}cccc@{}}
\toprule
\textbf{Regime} & \textbf{Selected noise} & \multicolumn{3}{c}{\textbf{\rev{Coverage}}} \\
\cmidrule(lr){3-5}
 & \textbf{$75\to600$} & \textbf{\rev{Signals}} & \textbf{\rev{Noise}} & \textbf{\rev{All selected}} \\
 & & \textbf{$75\to600$} & \textbf{$75\to600$} & \textbf{$75\to600$} \\
\midrule
Low-dimensional baseline & $2.73\to3.71$ & $0.906\to0.905$ & $0.845\to0.851$ & $0.881\to0.879$ \\
Moderate candidate set & $4.50\to6.76$ & $0.928\to0.892$ & $0.903\to0.815$ & $0.915\to0.844$ \\
Correlated predictors & $4.20\to6.20$ & $0.928\to0.901$ & $0.914\to0.852$ & $0.921\to0.871$ \\
Larger candidate set & $6.37\to11.10$ & $0.982\to0.871$ & $0.941\to0.772$ & $0.956\to0.798$ \\
Larger candidate set, correlated & $5.24\to8.94$ & $0.962\to0.897$ & $0.942\to0.841$ & $0.951\to0.859$ \\
\rev{Weak signals (correlated)} & $3.81\to6.34$ & $0.945\to0.906$ & $0.924\to0.811$ & $0.934\to0.847$ \\
Heavy censoring & $5.07\to8.26$ & $0.991\to0.902$ & $0.997\to0.845$ & $0.994\to0.864$ \\
Proxy/competing predictors & $4.49\to6.44$ & $0.948\to0.911$ & $0.936\to0.872$ & $0.939\to0.881$ \\
\bottomrule
\end{tabular}
\end{table}

The \emph{larger candidate set} regime (\(p=50\), \(\rho=0\)) illustrates this distinction particularly clearly. The mean number of selected noise predictors increased from 6.37 to 11.10, while coverage among selected noise coordinates decreased from 0.941 to 0.772. Coverage among selected signal coordinates remained higher but also decreased from 0.982 to 0.871. The decline in overall selected-coordinate coverage therefore reflects both a changing composition of the selected predictor set and deterioration in coverage within some coordinate groups. Similar, though less pronounced, patterns occurred under correlation, weak signals, and heavier censoring. Full setting-specific composition summaries are provided in the Supporting Material, Section S3 (Additional simulation results and sensitivities).

The heavy-censoring regime also illustrates why coverage and interval width should be considered jointly. At \(n=75\), re-selection with bootstrap CV-min tuning was strongly conservative, with coverage 0.994 and mean interval width 2.440. Re-selection at \rev{fixed~\(\widehat\lambda_{\min}\)} had coverage 0.915 with mean width 1.167. By \(n=600\), coverage under re-selection with bootstrap CV-min tuning was 0.864, with coverage 0.902 among selected signal coordinates and 0.845 among selected noise coordinates. The small-sample overcoverage therefore coincides with very wide intervals and should not be interpreted as better inferential performance.

\begin{figure}[ht]
\centering
\includegraphics[width=0.82\linewidth]{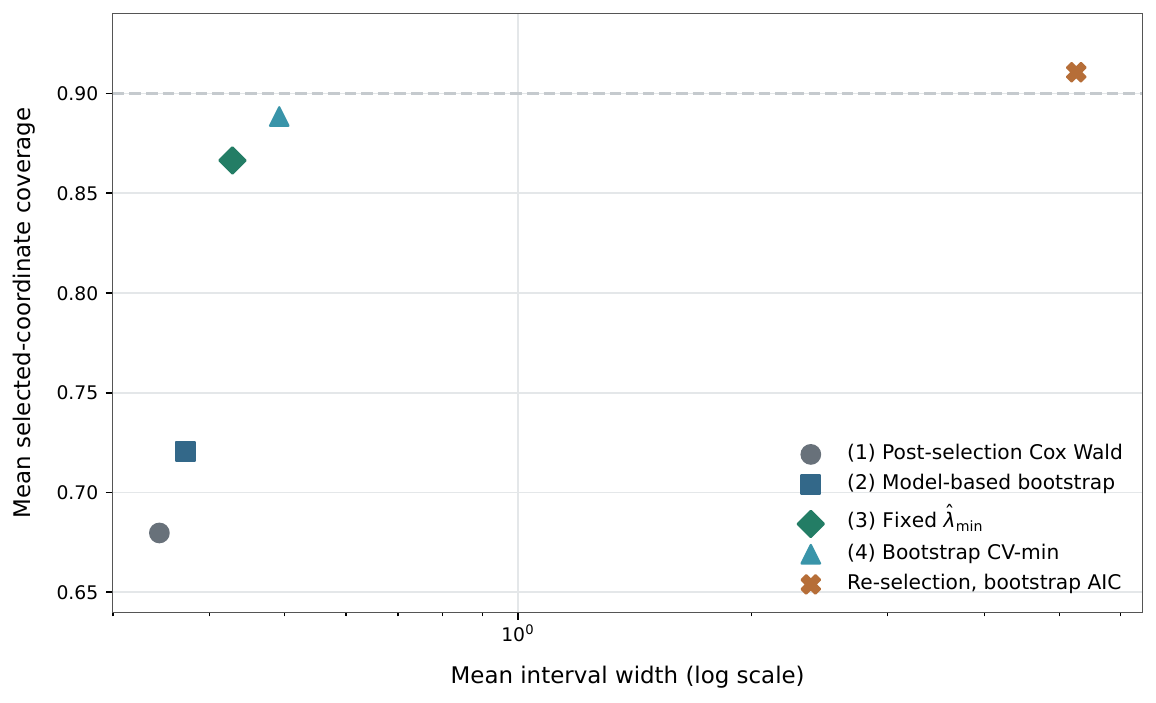}
\caption{Mean selected-coordinate coverage versus mean interval width for the main simulation comparison and the AIC tuning sensitivity. Each point summarises the 32 simulation settings and should therefore be interpreted as an overall descriptive comparison rather than as coefficient-specific performance. The width axis is logarithmic because AIC-based tuning produces substantially wider intervals. Bootstrap procedures use bootstrap-SE Wald intervals.}
\label{fig:coverage-width-tradeoff}
\end{figure}

The overall coverage--width relationship is shown in Figure~\ref{fig:coverage-width-tradeoff}. The AIC sensitivity is included to illustrate why coverage alone can be misleading, as its near-nominal mean coverage is accompanied by substantially larger intervals. \rev{For bootstrap-SE Wald intervals, the AIC sensitivity had a mean width of 5.257, compared with 0.492 for re-selection with bootstrap CV-min tuning (Figure~\ref{fig:coverage-width-tradeoff} and Supporting Material, Section~S3.2).}

\subsubsection{Selected-set recovery and bootstrap re-selection}
\label{sec:selection-accuracy-results}

\selrev{Table~\ref{tab:main-selection-composition} compares the original Cox--Lasso selection and the two bootstrap re-selection strategies. Across the 32 settings, the original CV-min selection recovered the complete active set $M_0$ in 2.10\% of Monte Carlo replicates, with mean Jaccard overlap 0.444. On average, it selected 3.726 active and 5.884 inactive predictors. As sample size increased, fewer active predictors were missed, but more inactive predictors were selected (Supporting Material, Section~S5). The complete active set was therefore recovered less often despite the improved detection of individual signals.}

\begin{table}[htbp]
\centering
{\color{selectiongreen}
\caption{\selrev{Numbers of active and inactive predictors selected by the original Cox--Lasso and the two bootstrap re-selection strategies. Values are equally weighted means across the 32 settings, with 900 Monte Carlo replicates per setting. Bootstrap entries are expected counts calculated from selection frequencies over $B=500$ bootstrap samples per original replicate. They describe selection under the bootstrap procedure, not recovery probabilities under the original simulation design.}}
\label{tab:main-selection-composition}
\small
\setlength{\tabcolsep}{5pt}
\begin{tabular}{@{}lrrrr@{}}
\toprule
\textbf{Selection procedure} & \textbf{Mean size} & \textbf{TP} & \textbf{FP} & \textbf{FN} \\
\midrule
Original CV-min & 9.610 & 3.726 & 5.884 & 0.024 \\
Re-selection, fixed $\widehat\lambda_{\min}$ & 11.873 & 3.694 & 8.179 & 0.056 \\
Re-selection, bootstrap CV-min & 13.578 & 3.696 & 9.882 & 0.054 \\
\bottomrule
\end{tabular}
}
\end{table}

\selrev{Compared with re-selection at fixed $\widehat\lambda_{\min}$, reapplying CV-min within each bootstrap sample increased the expected number of falsely selected predictors by 1.703 (paired Monte Carlo standard error 0.0088; approximate 95\% Monte Carlo interval 1.686 to 1.720). The corresponding increase in correctly selected active predictors was only 0.0026. The excess number of false positives increased from 1.439 at $n=75$ to 2.031 at $n=600$. Likewise, the expected Hamming distance to the original selected set increased from 6.684 under fixed-$\widehat\lambda_{\min}$ re-selection to 8.095 under bootstrap CV-min. These differences describe which predictors are selected in bootstrap samples; they do not, on their own, establish differences in interval coverage. Monte Carlo standard errors are based on comparisons within the same original simulated data sets, although the two procedures use separate bootstrap samples. Results by sample size, Monte Carlo standard errors, and the AIC sensitivity are reported in Supporting Material, Section~S5.}

\subsubsection{Coefficient-specific coverage}

The aggregate and signal/noise-stratified summaries combine information across different selected coefficients. To examine the heterogeneity that may be hidden by this aggregation, we additionally evaluate coverage separately for individual data-generating coefficients. For coefficient $\beta_j^\circ$, coverage is calculated among numerically available Monte Carlo replicates in which predictor $j$ was selected in the original sample. The target is the same fixed data-generating coefficient $\beta_j^\circ$ across these replicates, even when the other predictors selected alongside $j$ differ.

\begin{figure}[ht]
\centering
\includegraphics[width=0.99\linewidth]{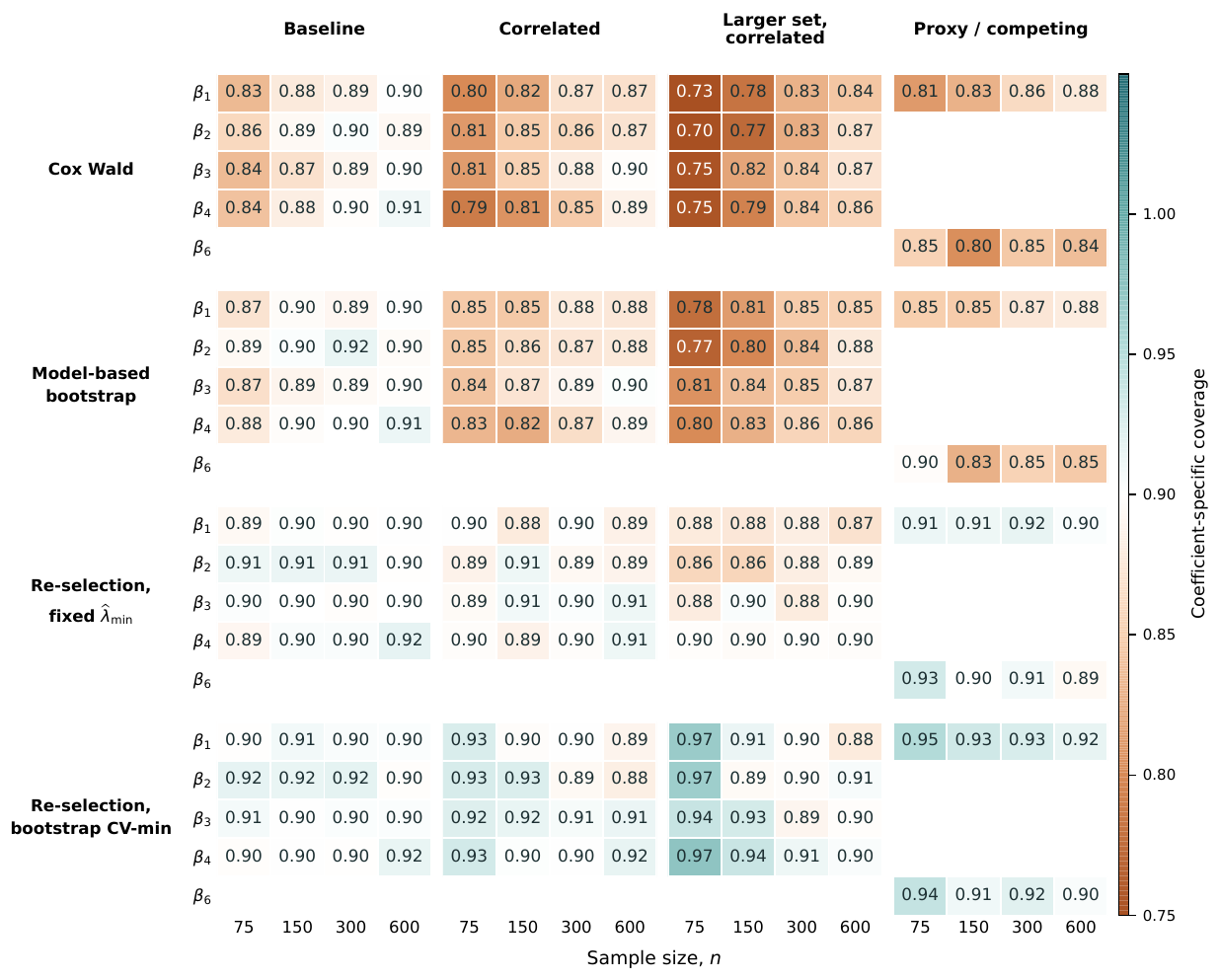}
\caption{Coefficient-specific coverage against the data-generating coefficients $\beta_j^\circ$ for originally selected predictors in four representative simulation regimes (nominal level 0.90). \rev{Rows show (1) post-selection Cox Wald, (2) model-based bootstrap without re-selection, (3) re-selection at fixed~$\widehat\lambda_{\min}$, and (4) re-selection with bootstrap CV-min tuning. Bootstrap procedures use bootstrap-SE Wald intervals. White corresponds to the nominal 0.90 coverage, whereas orange and blue-green indicate undercoverage and overcoverage respectively, with equal colour intensity for equal absolute deviations from 0.90 (up to saturation).} Only coefficients with nonzero data-generating values are displayed for each regime. Blank cells mark coefficients with zero data-generating values.}
\label{fig:coefficient-specific-coverage}
\end{figure}

Figure~\ref{fig:coefficient-specific-coverage} compares \rev{all four primary procedures, including re-selection at fixed~$\widehat\lambda_{\min}$,} for four representative regimes. For readability, only coefficients with nonzero data-generating values in the respective regime are displayed. The bootstrap procedures use bootstrap-SE Wald intervals. \rev{Coverage of the originally selected noise coefficients is reported separately in Supporting Material, Section~S3.4.}

The coefficient-specific results show variation that is hidden by the overall averages. For example, in the correlated larger candidate set at \(n=75\), coverage for \(\beta_1^\circ\) was 0.73 under post-selection Cox Wald and 0.97 under re-selection with bootstrap CV-min. At \(n=600\), the corresponding values were 0.84 and 0.88. Re-selection therefore often brought coverage closer to 0.90, but could also produce overcoverage, particularly at smaller sample sizes. \rev{Undercoverage means that intervals exclude the target more often than the nominal noncoverage rate and is especially concerning for nominal error control. Overcoverage is conservative in this respect but may be associated with wider, less informative intervals. Thus, a symmetric colour scale around 0.90 visualises deviation from the nominal level without assigning equal inferential consequences to the two directions.} The corresponding coefficient-specific analysis against the Cox projection targets is given in Supporting Material, Section~S3.4. \rev{These plots use a diverging colour scale centred at the nominal level, so neither undercoverage nor overcoverage is visually presented as inherently favourable.}

\subsubsection{Interval-construction sensitivity}

The primary comparison uses bootstrap-SE Wald intervals and evaluates coverage against the data-generating coefficients. \rev{Supporting Material, Section~S3.2, also evaluates percentile intervals against both the data-generating coefficients and the Cox projection coefficients for the selected model. These analyses are supplementary to the primary bootstrap-SE Wald comparison.}

\section{NKI70 breast-cancer application}
\label{sec:application-plan}

The NKI70 application illustrates selected-set and coefficient-level instability in an applied Cox--Lasso analysis. The focus is methodological, not biomarker discovery; genes entering only through bootstrap re-selection are used as stability diagnostics.

\subsection{Data and analysis}

We use the NKI70 breast cancer data set, distributed as \texttt{nki70} in the R package \texttt{penalized}, with a time-to-event outcome and a 70-gene expression signature \citep{vandeVijver2002,penalizedNKI70}. The analysed data contain 144 patients, 70 gene-expression variables, and 48 observed events, defined as metastasis or death. Clinical variables contained in the data object are excluded. Gene expression is standardised, and \rev{10-fold} cross-validation with the CV-min tuning rule is used to select \(\widehat\lambda_{\min}\). An unpenalised Cox model is then refitted using the selected predictor set. The observed tuning-parameter value is \(\widehat\lambda_{\min}=0.0983\), yielding four selected genes: QSCN6L1, ZNF533, IGFBP5.1, and PRC1.

The application uses the same semiparametric model-based bootstrap as described in Section~\ref{sec:bootstrap}, with \(B=500\) replicates. We compare the model-based bootstrap without re-selection, re-selection at \rev{fixed~\(\widehat\lambda_{\min}\)}, re-selection with bootstrap CV-min tuning, and AIC-based tuning as a sensitivity analysis. For predictor-level summaries, we use the same coefficient representation as in the simulation study, with zero denoting non-selection in a bootstrap sample rather than a fitted zero coefficient.

\subsection{Results}

\selrev{The true active gene set is unknown in NKI70, so we cannot determine TP, FP, or FN. We instead examine how often the original four-gene set and its individual genes are selected in the bootstrap samples.}

\begin{table}[ht]
\centering
\caption{Selected-set stability diagnostics for the NKI70 analysis based on \(B=500\) bootstrap replicates. ``Exact set'' is the proportion reproducing the observed predictor set; ``Mean Jaccard'' is mean overlap with that set; ``Mean no. selected'' and ``SD no. selected'' summarise the size of the bootstrap-selected set; \rev{``Refit retained'' is the proportion of bootstrap samples with a coefficient vector from the unpenalised Cox refit retained for interval calculation.}}
\scriptsize
\setlength{\tabcolsep}{3.2pt}
\resizebox{\linewidth}{!}{%
\begin{tabular}{lccccc}
\toprule
\textbf{Method} & \textbf{Exact set} & \textbf{Mean Jaccard} & \textbf{Mean no. selected} & \textbf{SD no. selected} & \textbf{\rev{Refit retained}} \\
\midrule
Model-based bootstrap & 1.000 & 1.000 & 4.000 & 0.000 & 1.000\\
Re-selection, \rev{fixed~$\widehat\lambda_{\min}$} & 0.006 & 0.335 & 5.880 & 2.064 & 1.000\\
Re-selection, bootstrap CV-min & 0.000 & 0.256 & 9.238 & 4.365 & 1.000\\
Re-selection, bootstrap AIC tuning & 0.000 & 0.104 & 51.522 & 23.476 & 0.930\\
\bottomrule
\end{tabular}%
}
\label{tab:nki70-support}
\end{table}

For the model-based bootstrap without re-selection, exact-set reproduction and Jaccard overlap equal one by construction, not because variable selection was empirically stable. In contrast, Table~\ref{tab:nki70-support} shows substantial instability under re-selection. At \rev{fixed~\(\widehat\lambda_{\min}\)}, the observed four-gene predictor set is reproduced in only 0.6\% of bootstrap samples, with a mean Jaccard overlap of 0.335 and 5.88 genes selected on average. Bootstrap CV-min tuning further reduces the mean Jaccard overlap to 0.256 and increases the mean selected-set size to 9.24 genes. \selrev{As in the simulations, bootstrap CV-min selects more predictors on average (Section~\ref{sec:selection-accuracy-results}). In NKI70, however, we cannot determine how many of the additional predictors are truly inactive.}

AIC-based tuning produces substantially larger selected predictor sets, with 51.52 genes selected on average (standard deviation 23.48). \rev{For 465 of the 500 bootstrap samples (93\%), an unpenalised Cox coefficient vector was available for interval calculation; the remaining 35 samples did not contribute a coefficient vector. This numerical availability does not assess selection accuracy or inferential validity.} Because this tuning rule differs from the CV-min rule used in the observed data and leads to very large selected predictor sets relative to the 48 observed events, we regard it as an extreme tuning sensitivity rather than as a primary analysis.

Table~\ref{tab:nki70-coefficients} reports the observed coefficient estimates and 90\% intervals for the four genes selected in the observed data. Reverse-percentile basic intervals are reported as an interval-construction sensitivity in the Supporting Material, \rev{Section S4.1, Table S9}.

\begin{table}[ht]
\centering
\caption{NKI70 unpenalised Cox coefficient estimates and nominal 90\% confidence intervals for the four genes selected in the observed data. Bootstrap procedures use bootstrap-SE Wald intervals based on \(B=500\) replicates.}
\scriptsize
\setlength{\tabcolsep}{3pt}
\resizebox{\linewidth}{!}{%
\begin{tabular}{lccccc}
\toprule
\textbf{Gene} & \textbf{Estimate} & \textbf{Post-selection Cox Wald} & \textbf{Model-based bootstrap} & \textbf{Re-selection, \rev{fixed~$\widehat\lambda_{\min}$}} & \textbf{Re-selection, bootstrap CV-min}\\
\midrule
QSCN6L1 & 0.200 & [-0.051, 0.451] & [-0.071, 0.471] & [-0.034, 0.434] & [-0.068, 0.468]\\
ZNF533 & -0.288 & [-0.565, -0.012] & [-0.584, 0.007] & [-0.643, 0.066] & [-0.684, 0.107]\\
IGFBP5.1 & 0.319 & [0.121, 0.517] & [0.080, 0.558] & [-0.086, 0.724] & [-0.123, 0.761]\\
PRC1 & 0.578 & [0.268, 0.888] & [0.257, 0.899] & [0.129, 1.027] & [0.079, 1.077]\\
\bottomrule
\end{tabular}%
}
\label{tab:nki70-coefficients}
\end{table}

\begin{figure}[ht]
\centering
\includegraphics[width=0.88\linewidth]{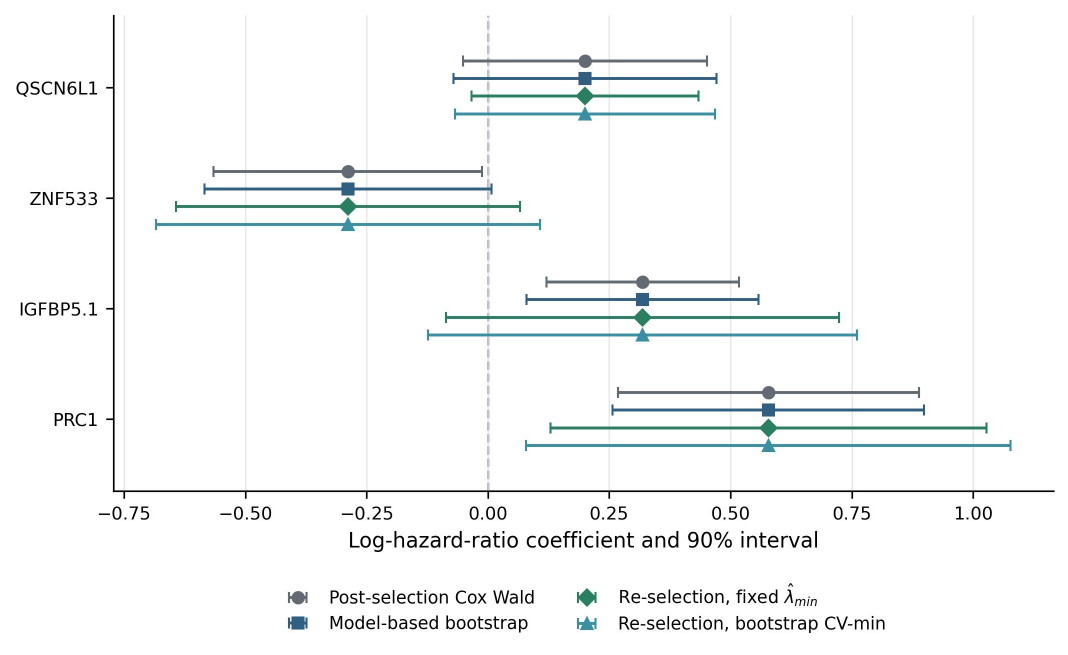}
\caption{NKI70 coefficient estimates and 90\% intervals for the four genes selected in the observed data. Bootstrap procedures use bootstrap-SE Wald intervals.}
\label{fig:nki70-ci}
\end{figure}

Figure~\ref{fig:nki70-ci} shows the same intervals graphically.

The coefficient-level consequences of re-selection differ across genes. PRC1 remains separated from zero under the model-based bootstrap and both re-selection procedures. IGFBP5.1 remains separated from zero under the model-based bootstrap, but both re-selection intervals include zero. QSCN6L1 and ZNF533 include zero under all three bootstrap procedures. These differences do not by themselves imply that re-selection removes genuine signals: including zero can instead reflect the additional uncertainty associated with changes in the selected predictor set. Conversely, an interval that excludes zero when selection is held fixed may understate that uncertainty. These are descriptive comparisons, not validity guarantees for any particular gene.

The original genes also differ in coordinate-level stability: PRC1 is re-selected most frequently under both re-selection rules, whereas QSCN6L1 is re-selected least frequently. Detailed re-selection and zero frequencies are given in Supporting Material, Section S4.2.

The reverse-percentile basic sensitivity analysis \rev{(Supporting Material, Section S4.1, Table S9)} gives the same qualitative conclusion for PRC1 but differs for QSCN6L1 and IGFBP5.1 under re-selection at \rev{fixed~\(\widehat\lambda_{\min}\)}, where the reverse-percentile basic intervals narrowly exclude zero while the bootstrap-SE Wald intervals include zero. We therefore retain the common bootstrap-SE Wald construction for the main comparison.

\begin{figure}[ht]
\centering
\includegraphics[width=0.72\linewidth]{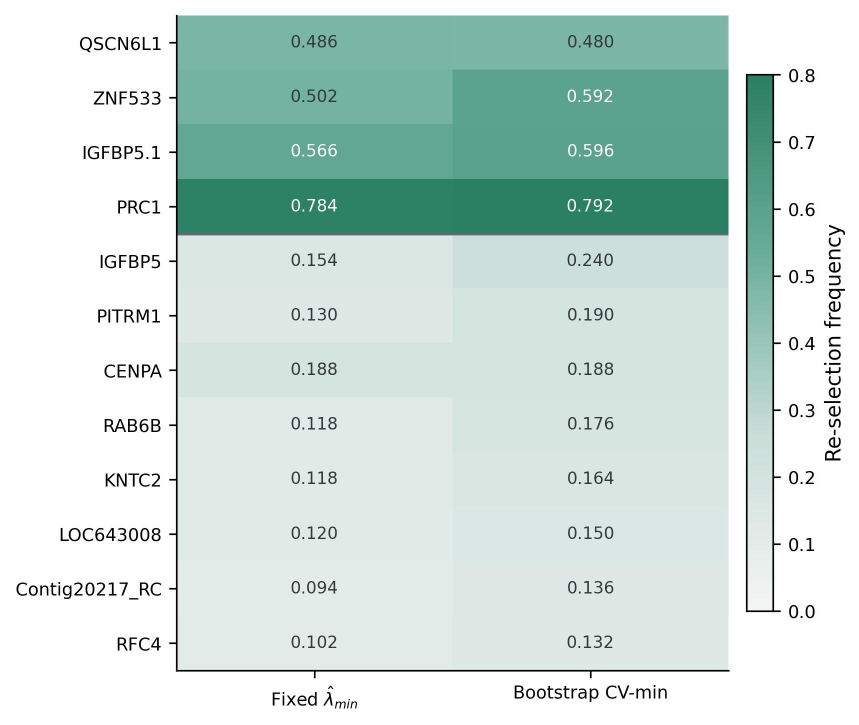}
\caption{NKI70 re-selection frequencies under \rev{fixed~\(\widehat\lambda_{\min}\)} and bootstrap CV-min tuning. The first four genes were selected in the observed analysis. The remaining rows are genes that entered only under bootstrap re-selection.}
\label{fig:nki70-selection-frequency}
\end{figure}

Re-selection also identifies genes that were not selected in the observed fit but enter repeatedly in bootstrap samples. \rev{Figure~\ref{fig:nki70-selection-frequency} compares re-selection frequencies for the four originally selected genes and eight frequently selected additional genes. Among these eight bootstrap-emerging genes, frequencies were higher under bootstrap CV-min tuning for seven and identical under both strategies for CENPA. This descriptive pattern is consistent with the larger bootstrap-selected sets under CV-min tuning in Table~\ref{tab:nki70-support}, but does not imply that re-tuning always increases the selection frequency of an individual gene.}

The bootstrap-emerging predictors also indicate possible proxy substitution. IGFBP5 is almost perfectly correlated with the originally selected IGFBP5.1 (\(|r|=0.978\)), PRC1 is strongly correlated with NUSAP1 (\(|r|=0.830\)) and CENPA (\(|r|=0.818\)), and QSCN6L1 is moderately correlated with CDCA7 (\(|r|=0.625\)). These relationships provide a plausible explanation for some changes in the selected predictor set under re-selection. \rev{The near-collinearity of IGFBP5 and IGFBP5.1 makes their separate selection difficult to interpret. Pre-specified correlation filtering could reduce such redundancy, but would change the candidate predictor set and was not applied in this methodological illustration.} Additional NKI70 interval and proxy diagnostics are provided in the Supporting Material, Section S4 (Additional NKI70 analyses).

Overall, the NKI70 analysis shows substantial selected-set instability while the consequences for individual reported coefficients differ. In particular, across the main interval constructions, the interval for PRC1 continues to exclude zero, unlike the re-selection intervals for the other originally selected genes. This pattern is descriptive and does not establish that the corresponding association is stable in the target population.

\FloatBarrier

\section{Discussion}

This study examines whether uncertainty assessment after Cox--Lasso selection should reflect instability of the selected predictor set. Our primary simulation evaluation concerns the known data-generating coefficients of originally selected predictors. Repeating selection within the model-based bootstrap generally moved coverage closer to 0.90 than procedures that held the selected set fixed. Intervals also became wider, and coverage remained variable across settings and coefficients. Thus, re-selection can capture uncertainty omitted by fixed-set procedures, but does not guarantee nominal coverage.

Coverage alone does not describe the performance of a procedure after selection. Interval width, the composition and stability of selected sets, and results for individual coefficients provide complementary information. The AIC sensitivity, for example, showed near-nominal average coverage but much wider intervals. The supplementary analysis using Cox projection targets also showed that the interval construction matters. Thus, the selection strategy and the choice of interval are distinct parts of the uncertainty assessment.

\rev{Model averaging provides a complementary approach to model uncertainty \citep{HjortClaeskens2003,HjortClaeskens2006}. While it combines estimates across candidate models, our procedure retains the coefficient reported from the originally selected unpenalised Cox refit and uses bootstrap re-selection to assess its uncertainty. The approaches therefore need not address the same inferential target or reporting objective. We did not include model averaging in the simulation comparison because the study was designed to isolate the contribution of repeated Cox--Lasso selection to interval uncertainty for the original refit. A direct comparison would require specifying a common estimand and a corresponding model-averaging procedure. Our findings should not be interpreted as evidence that re-selection outperforms model averaging.}

\selrev{Across the simulation settings, reapplying CV-min during bootstrap re-selection increased the average number of selected inactive predictors, while the number of selected active predictors changed very little. These results complement the coverage and interval-width comparisons, but do not establish whether either procedure achieves nominal coverage.}

The NKI70 application illustrates why selected-set and coefficient-level stability should be distinguished. In particular, the wider re-selection intervals for IGFBP5.1 should not automatically be read as obscuring an established effect: they also show how much the inference can change when uncertainty in the preceding selection step is represented. The complete selected predictor set was rarely reproduced under re-selection, whereas the consequences for the four originally selected genes differed substantially. Correlated gene-expression variables also appeared to substitute for one another during selection. Thus, instability of the selected model as a whole need not imply the same degree of instability for every reported coefficient, and selected-set and coordinate-level diagnostics provide complementary information.

Coverage and interval width quantify different aspects of inference, while re-selection frequencies, selected-set overlap, and selected-set composition describe model stability. We report these measures separately because no single one characterises the full procedure. \rev{The archived simulation outputs retain marginal bootstrap selection frequencies but not the complete predictor sets from individual bootstrap draws. Consequently, expected selection counts and Hamming distances can be assessed, whereas bootstrap exact-set agreement and Jaccard overlap with the originally selected or true active set cannot be calculated from these outputs. Retaining complete bootstrap-selected sets in an additional analysis would permit these joint stability measures to be examined alongside interval coverage and width; such results are not part of the present study.} Developing a principled joint assessment of interval performance and selection stability remains an open direction for further work.

Several limitations define the scope of these findings. The bootstrap is model-based and conditional on the observed covariate vectors, so its operating characteristics depend on the fitted time-to-event and censoring models. The plug-in generating law is not necessarily the true time-to-event law, especially if the selected refit omits relevant predictors. The NKI70 stability frequencies should not be interpreted as nonparametric estimates of population-wide selection probabilities. The simulation design is deliberately focused, with \(p\le50\) and a limited range of signal, correlation, and censoring structures, and therefore does not cover ultra-high-dimensional settings or all forms of model misspecification. CV-min is the primary tuning rule. Apart from the exploratory AIC sensitivity, other tuning rules and regularisation methods are not systematically investigated. The primary coverage summaries also concern originally selected predictors and do not assess inference for signals omitted by the original selection. Under re-selection, a coefficient is stored as zero when its predictor is not selected, so the resulting distribution is tied to the specified selection and refitting procedure. Finally, the study provides finite-sample evidence but does not establish general post-selection validity for this non-smooth procedure.

\rev{For the four primary procedures, evaluating the same intervals against the Cox projection targets instead of the data-generating coefficients changed mean coverage by less than 0.001 (Supporting Material, Section~S3.5).} Some individual settings and coefficient groups showed larger differences. The similar overall results do not imply that the two targets coincide or that re-selection corrects bias due to omitted predictors. Our findings support re-selection as an empirical way to reflect selection variability in these settings, not as a general correction for post-selection inference.

\section*{Supplementary materials}
The accompanying Supporting Material provides detailed estimands, additional simulation summaries,
selection diagnostics, sensitivity analyses and supporting results for the NKI70 application.
The Supporting Material is available as a separate ancillary PDF accompanying this preprint.

\section*{Declaration of conflicting interests}

The authors declare no competing interests.

\section*{Ethical considerations}
The simulation component uses synthetically generated data. The applied component is a secondary
methodological analysis of the previously published, publicly distributed \texttt{nki70} data set
in the R package \texttt{penalized} \citep{vandeVijver2002,penalizedNKI70}. No new participants were
recruited and no new patient-level data were collected for the present study. The original
clinical study is cited as the source of the data.

\section*{Consent to participate and for publication}
The present simulation and secondary methodological analysis did not involve recruitment or contact
with patients. No directly identifying participant information is reported in this article.

\section*{Funding}

This work was supported by the Deutsche Forschungsgemeinschaft (DFG, German Research Foundation), project number 439942859. \rev{Moreover, Andreas Groll gratefully acknowledges financial support from the Deutsche Forschungsgemeinschaft (DFG) within TRR~391: Spatio-temporal Statistics for the Transition of Energy and Transport (project number 520388526).}

\section*{Data availability}

The simulation data are generated from the analysis scripts and require no external data source. The NKI70 data example uses the publicly available \texttt{nki70} data set distributed with the R package \texttt{penalized}. This data set contains metastasis-free follow-up time, event status, clinical variables, and the 70-gene expression signature for 144 lymph-node-positive breast cancer patients \citep{vandeVijver2002,penalizedNKI70}. In the data example, we use the outcome variables \texttt{time} and \texttt{event} and the 70 gene-expression columns only; clinical covariates are not included in the genes-only analysis. The reproducibility release will include the random seeds, design files, collected result summaries, and the complete preprocessing and analysis scripts.

\section*{Code availability}

\selrev{A maintained implementation of the methods is available in the R package \texttt{selSurv} (version 0.1.1; \url{https://cran.r-project.org/package=selSurv}).} In addition, a study-specific GitHub repository, \path{cox-lasso-bootstrap-reselection-study}, will archive the exact production simulation and NKI70 scripts, simulation design, compact audited outputs, random seeds, software-version information, and manuscript sources used for this study. 

\section*{Author contributions}

L.S. conceived the study, implemented all methods, performed the analyses, and wrote the manuscript. S.F.-W. and A.G. contributed to the methodological development and provided supervision, peer review, and critical feedback. All authors read and approved the final manuscript.

\section*{Declaration of generative AI and AI-assisted technologies in the manuscript preparation process}
During the preparation of this work, the authors used AI-assisted tools, including ChatGPT (OpenAI), Claude Code (Anthropic), and Codex (OpenAI), to assist with language and manuscript organization, and with the review, debugging, and execution support of computational workflows. The authors reviewed the AI-assisted material, independently assessed the scientific methods and reported results, and take full responsibility for the article's content and conclusions.

\end{document}